\documentclass[runningheads]{llncs}
\usepackage{graphicx}
\usepackage{booktabs}
\usepackage{tikz}
\usetikzlibrary{arrows.meta,positioning}

\begin{document}
\title{Trustworthy AI Teammates for Software Engineering: A Research Agenda for Lifecycle Context, Explainability, and Agentic Collaboration}
\titlerunning{Trustworthy AI Teammates for Software Engineering}
%
\author{Glaucia Melo\inst{1}\orcidID{0000-0003-0092-2171} 
}
\authorrunning{G. Melo}
%
\institute{Toronto Metropolitan University, Toronto, ON M5B2K3, Canada 
\email{glaucia@torontomu.ca}\\
}
\maketitle              
\begin{abstract}
Large Language Models are rapidly changing software engineering (SE) practice by supporting code generation, testing, debugging, documentation, and program comprehension. Yet most current AI coding assistants remain reactive tools: they respond to local prompts, rely on limited context, and provide suggestions whose rationale is often difficult for developers to inspect. This paper argues that the next phase of AI-assisted SE requires a shift from coding assistants to trustworthy AI teammates. We present a research agenda for human-AI co-evolution in SE, organized around three requirements. First, AI systems must become lifecycle-aware, reasoning across requirements and other deployment artifacts rather than isolated files or prompts. Second, they must provide principled explanations that expose assumptions, evidence, trade-offs, and risks in ways that support developer judgment. Third, they must enable agentic collaboration, in which specialised AI agents participate in software workflows while preserving human oversight and accountability. We propose an evaluation framework that moves beyond isolated functional correctness toward collaborative quality, including contextual alignment, trust calibration, maintenance recovery cost, and traceability. We conclude by outlining open challenges for designing AI systems that improve software quality without weakening human understanding, agency, or responsibility.

\keywords{AI-assisted software engineering \and large language models \and human-AI collaboration \and trustworthy AI \and software maintainability \and explainable AI \and agentic systems \and evaluation framework.}
\end{abstract}

\section{Introduction}

Artificial Intelligence (AI) has moved from the margins of software engineering (SE) into everyday development practice, reshaping how developers write, review, test, and maintain software~\cite{alenezi2025,durrani2024}. Large Language Models (LLMs) and AI coding assistants are now used to generate code, write tests, summarize documentation, explain APIs, and support debugging tasks~\cite{chen2021,roziere2023}. The appeal is clear: these tools can reduce friction in routine programming work, accelerate prototyping, and lower the barrier to entry for developers working in unfamiliar languages, frameworks, or codebases~\cite{melo2025quality,hamza2024}.

However, the rapid adoption of AI coding assistants has also exposed a deeper problem. Current systems are powerful, but they are still largely reactive. A developer asks a question, provides a prompt, or opens a file; the assistant responds. This interaction model works well for localized coding tasks, but it does not reflect how software systems are actually built, maintained, and governed~\cite{treude2025,jin2024}. Software development is not only code production. It involves requirements negotiation, architectural decisions, design constraints, testing, deployment, maintenance, team communication, and long-term accountability. When AI-generated code enters this lifecycle, it becomes part of a socio-technical system that future developers must understand, modify, and trust~\cite{lo2023,pezze2025}.

This creates a shift in the central research question. The field has spent considerable effort asking whether LLMs can generate correct code~\cite{chen2021,roziere2023}. That question remains important, but it is no longer sufficient. We must also ask whether AI-assisted software can remain understandable, reliable, secure, and maintainable over time. Prior work has shown that LLM-generated code may contain functional errors, security weaknesses, or maintainability problems even when it appears plausible or passes initial tests~\cite{siddiq2023,liu2023your,melo2025quality,jamil2025can}. A code suggestion may pass visible tests and still introduce hidden maintenance debt, violate project conventions, obscure design rationale, or reduce a developer's understanding of the system~\cite{dillmann2024,hamza2024,welter2025}.

In this paper, we argue that software engineering needs to move beyond the metaphor of AI as a coding tool and toward a more demanding vision: \textit{AI as a trustworthy teammate}. We use the term \textit{human-AI co-evolution} to describe a software engineering future in which developers, teams, tools, and AI systems adapt to one another over time~\cite{dubey2020,lo2023,fischer2023}. This does not mean replacing developers or granting unrestricted autonomy to AI systems. Rather, it means designing AI systems that can participate in development workflows while remaining transparent, contestable, auditable, and aligned with human goals~\cite{ahmad2023,bohme2025}.

This vision is especially important for global and resource-diverse software ecosystems. AI assistants may expand access to software development knowledge, support education, and help teams work across languages, regions, and levels of experience~\cite{alenezi2025,durrani2024}. At the same time, they may also reproduce inequities if they are evaluated only on narrow benchmarks, trained primarily on dominant programming cultures, or deployed without attention to trust, accountability, and local context~\cite{bohme2025,Pearce2025}. A human-centered research agenda is therefore necessary not only for technical quality, but also for responsible adoption. As a forward-looking agenda, this paper combines established findings with emerging, and in some cases early-stage, evidence; we therefore present the requirements and evaluation metrics that follow as testable proposals rather than validated results. This paper makes three contributions:

\begin{itemize}
    \item We identify key limitations of current AI-assisted software engineering workflows, focusing on local context, opaque rationale, and weak support for long-term maintainability.
    \item We propose three requirements for trustworthy AI teammates in SE: lifecycle-aware context, principled explainability, and agentic collaboration.
    \item We outline an evaluation framework and research agenda for studying human-AI co-evolution in software engineering, including metrics for contextual alignment and trust calibration.
\end{itemize}

\section{Why Current AI Assistance Falls Short}

The first wave of LLM-based software engineering tools has shown that AI can be useful in development practice. These systems can produce compilable code, generate test cases, summarize unfamiliar APIs, and help developers overcome routine implementation barriers~\cite{chen2021,roziere2023,treude2025}. Yet their usefulness should not be mistaken for maturity. The dominant interaction model still treats AI assistance as a localized response mechanism rather than as a participant in the broader software lifecycle~\cite{lo2023,pezze2025}.

\subsection{Local Context Is Not Lifecycle Context}

Most AI coding assistants operate over limited context: the current prompt, the open file, nearby code, or a retrieved set of snippets~\cite{jin2024,elsisi2025}. This can be enough for small tasks, but it is often insufficient for software engineering work that depends on architectural constraints, historical decisions, issue discussions, test failures, deployment environments, or team conventions~\cite{fischer2023,durrani2024}. A generated patch may be syntactically correct and even pass visible tests, while still violating an implicit design decision or duplicating logic that the project has intentionally centralized elsewhere.

The problem is not simply that context windows are too small. The deeper issue is that current tools rarely distinguish between different kinds of context or explain why a particular piece of context was used. Requirements, bug reports, commit histories, code review comments, API contracts, and CI logs have different evidentiary value. A trustworthy AI teammate should not merely retrieve more context; it should select, rank, and justify the context that informs its suggestions~\cite{lewis2020,barnett2024,caumartin2026}.

\subsection{Correct Output Is Not the Same as Trustworthy Output}

A second limitation concerns explanation. Developers are often asked to accept or reject AI-generated suggestions without a clear account of the assumptions behind them. In many cases, the assistant explains what the generated code does, but not why that solution fits the project, which alternatives were considered, or what trade-offs were made~\cite{ahmad2023,samir2023,agarwal2020}. This weakens developer agency. If developers cannot inspect the rationale behind a suggestion, they may either over-trust the system or reject useful assistance because the risk feels too high~\cite{patrick2002building,ayele2025developer,Wang2024}.

This issue is particularly important for security, reliability, and maintainability. Prior empirical work has shown that LLM-generated code can contain subtle vulnerabilities, functional errors, or maintainability concerns, even when the output appears plausible~\cite{siddiq2023,liu2023your,melo2025quality,jamil2025can}. The concern is not that AI-generated code is always worse than human-written code. Rather, the concern is that its failure modes can be difficult to detect, especially when the code is fluent, idiomatic, and produced with confidence~\cite{liu2024,zhang2023,toth2024}.

\subsection{Collaboration Requires Accountability}

A third limitation is that current AI-assisted workflows often lack clear accountability. When an AI assistant suggests a change, the human developer remains formally responsible for reviewing and integrating it. However, as AI systems become more agentic, this boundary becomes less straightforward~\cite{he2025,ferrag2025}. If an AI agent retrieves issue context, edits multiple files, updates tests, and proposes a pull request, then the development workflow begins to resemble collaboration rather than tool use.

Collaboration requires traceability. A human teammate can usually be asked why a change was made, what requirement it addresses, and what risks remain. AI systems need analogous mechanisms. Without logs, provenance records, rationales, and human checkpoints, agentic AI may increase productivity while weakening the team's ability to explain and govern its own software~\cite{ahmad2023,abule2025,zhang2023,bohme2025,Pearce2025}.

\section{From Coding Tools to AI Teammates}

We propose that the next phase of AI-assisted software engineering should be organized around the goal of building trustworthy AI teammates. This requires a shift from tools that generate code on demand to systems that participate in development workflows while preserving human control~\cite{hamza2024,dubey2020}. We identify three core requirements: lifecycle-aware context, principled explainability, and agentic collaboration.

\begin{figure}[t]
    \centering
    \includegraphics[width=0.7\linewidth]{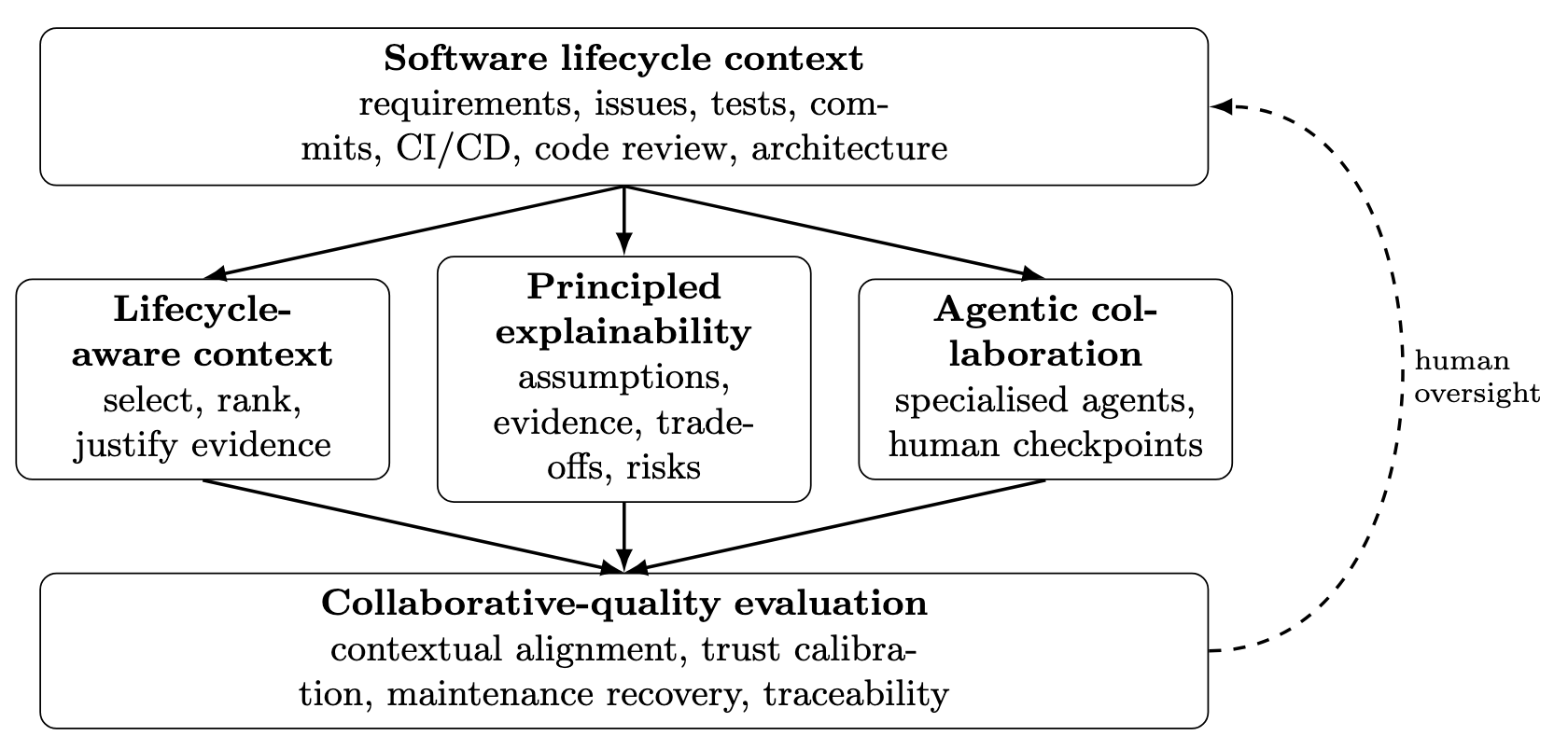}
    \caption{Overview of the proposed framework for trustworthy AI teammates in software engineering. Lifecycle context informs three requirements (lifecycle-aware context, principled explainability, and agentic collaboration) whose outputs are assessed along collaborative-quality evaluation dimensions. Human oversight closes the loop, supporting human--AI co-evolution over time.}
\label{fig:framework}
\end{figure}

These three requirements are not offered as an exhaustive taxonomy. They were derived by analyzing the recurring limitations set out in the previous section (local rather than lifecycle context, opaque rather than inspectable rationale, and tool-like rather than accountable collaboration) as they appear across the recent empirical literature on LLM-based software engineering~\cite{siddiq2023,liu2023your,melo2025quality,jamil2025can}. In this sense the framework is a conceptual synthesis of that literature and of current practice rather than the product of a formal systematic review. The framing is also deliberate. General frameworks for trustworthy AI emphasize model-level properties such as fairness, robustness, and transparency; the human--autonomy teaming literature studies interdependence largely outside software engineering, in domains such as aviation and command-and-control~\cite{oneill2022humanautonomy}; and work on multi-agent systems for software engineering has so far focused mainly on task performance~\cite{he2025,ferrag2025}. Our contribution is to specialize these strands to the software lifecycle, grounding each requirement in software engineering artifacts and decisions and pairing it with an evaluation dimension tailored to SE workflows. Figure~\ref{fig:framework} summarizes the resulting framework.

\subsection{Lifecycle-Aware Context}

A trustworthy AI teammate must reason across the software development lifecycle, not only across source code. Lifecycle-aware context includes the artifacts and signals that shape software decisions: requirements, issues, user stories, architecture documents, tests, dependency files, CI/CD logs, deployment traces, commit histories, code review discussions, and previous bug reports~\cite{fischer2023,durrani2024,elsisi2025}.

This does not mean that every artifact should be added indiscriminately to an AI prompt. Expanding the context an assistant consumes has real costs. Retrieving and encoding more artifacts increases token usage and monetary cost, adds latency, and can dilute the signals that matter with noise, so that more context does not reliably mean better suggestions. It also raises privacy and confidentiality concerns, because issue trackers, commit messages, and internal design documents may contain sensitive or regulated information that should not be sent to an external model. Finite context windows add a further limit: even relevant artifacts cannot all be included at once, so the system must choose. For these reasons, lifecycle-aware assistance requires mechanisms for context selection and context governance rather than indiscriminate retrieval. The AI system should be able to identify which artifacts are relevant to a task, explain why they matter, and distinguish between authoritative and weak signals~\cite{lewis2020,barnett2024}. For example, when proposing a bug fix, an AI teammate might cite the original issue report, the failing test, the previous commit that introduced the regression, and the architectural constraint that limits acceptable solutions~\cite{caumartin2026}. Similar principles appear in context-adaptive systems, where integrating heterogeneous signals can support more robust decision-making in complex environments~\cite{carvalho2025llm,carvalho2025llmLLMGrid}.

This moves AI assistance from code completion toward software reasoning. The goal is not only to produce an answer, but to preserve the relationship between the answer and the lifecycle evidence that supports it.

\subsection{Principled Explainability}

Trustworthy AI teammates must also provide explanations that are meaningful to software engineers. Generic explanations are not enough. Developers need rationales connected to SE concepts such as requirements, invariants, design patterns, dependencies, security principles, test coverage, performance constraints, and maintainability trade-offs~\cite{ahmad2023,samir2023,agarwal2020}.

Principled explainability means that an AI system should make its reasoning inspectable at the level where engineering decisions are made. For instance, instead of saying only that a function was refactored ``for readability,'' an AI teammate should explain which duplication was removed, which project convention was followed, which tests protect the behavior, and what risk remains. Similarly, when proposing a security patch, the system should connect the change to a vulnerability pattern, explain why the existing behavior is unsafe, and identify any assumptions that a developer must verify~\cite{johnson2025explainingsoftwarevulnerabilitieslarge,AlDebeyan2024vulnerability}.

This kind of explanation supports calibrated trust. The goal is not to make developers trust AI more. The goal is to help developers trust AI appropriately: accepting useful suggestions, questioning uncertain ones, and overriding outputs that conflict with project goals~\cite{hamza2024,dubey2020,patrick2002building}.

\subsection{Agentic Collaboration}

The third requirement is agentic collaboration. Modern software engineering tasks are often too complex for a single prompt-response interaction. Bug localization, feature implementation, migration, refactoring, and security repair may require planning, retrieval, code modification, testing, and review. Multi-agent systems offer one possible path forward, where specialized agents perform different roles within a coordinated workflow~\cite{he2025,ferrag2025}.

For example, one agent may summarize an issue report, another may retrieve relevant files, a third may propose a patch, and a fourth may critique the change against tests, requirements, and security constraints. Recent work on multi-agent systems for repository-level bug localization illustrates this direction by decomposing complex repository-level tasks into specialized agent responsibilities~\cite{caumartin2026}. This division of labor mirrors the collaborative dynamics of a human development team and can support more robust problem-solving when agents communicate, share intermediate results, and expose their reasoning to human reviewers~\cite{epperson2025,wang2025}.

However, agentic collaboration must be designed with human oversight in mind. The key question is not how to make AI agents autonomous, but how to make their work inspectable, interruptible, and accountable within human software teams~\cite{ahmad2023,abule2025,bohme2025}.

This oversight is not automatic, and several risks specific to agentic workflows must be managed. The first is over-automation: as agents take on more of a task, developers may disengage and approve their output with little scrutiny, following a well-documented pattern of automation bias and complacency in which operators under-check reliable-seeming automation and miss its errors~\cite{parasuraman2010complacency}. This dynamic has already been observed in AI-assisted programming, where developers with access to an AI assistant wrote less secure code yet were more confident that it was secure~\cite{perry2023insecure}. A second risk is correlated failure: when several agents share a base model, prompt, or retrieved context, their errors may be correlated rather than independent, so a critic agent can systematically miss the same mistakes a generator agent makes, and agreement among agents can give a false impression of reliability. A third risk is that human review becomes nominal rather than meaningful; a checkpoint that is formally present but too rushed or too poorly supported to constitute real control. For review to be meaningful, developers need to see why an agent acted, to interrupt a workflow, and to reverse agent actions, so that control tracks their reasons and the system's behaviour remains traceable to a responsible human~\cite{santonidesio2018meaningful}.

These risks sharpen the question of accountability. In current practice the human developer remains formally responsible for AI-assisted changes, but responsibility is fair and effective only when it is matched by control. If an agentic system spreads a task across many steps that a developer cannot realistically inspect, the developer may be left answerable for outcomes they had no meaningful way to prevent, becoming what Elish terms a \emph{moral crumple zone}, the nearest human who absorbs blame when the wider socio-technical system fails~\cite{elish2019moral}. Avoiding this is a design requirement rather than an ethical afterthought: provenance records, rationale logging, and well-placed human checkpoints are what make it possible to attribute responsibility across the developer, the tooling, and the team in a way that is defensible after the fact.

\section{An Evaluation Framework for AI Teammates}

Existing code generation benchmarks are useful, but they are not sufficient for evaluating AI teammates. Benchmarks such as HumanEval focus primarily on isolated functional correctness: given a prompt, does the model produce code that passes tests?~\cite{chen2021} This is an important measure, but it does not capture whether the AI system understands project context, provides useful rationale, preserves maintainability, or supports human decision-making~\cite{nascimento2025,dillmann2024}.

We propose an evaluation framework that shifts attention from isolated output quality to collaborative quality. Table~\ref{tab:evaluation-framework} summarizes the relationship between current limitations, co-evolution requirements, and possible evaluation metrics.

\begin{table*}[t]
\centering
\caption{Evaluation dimensions for trustworthy AI teammates in software engineering.}
\label{tab:evaluation-framework}
\scriptsize
\begin{tabular}{p{0.28\textwidth} p{0.32\textwidth} p{0.30\textwidth}}
\toprule
\textbf{Current limitation} & \textbf{Requirement for AI teammates} & \textbf{Possible evaluation metric} \\
\midrule
AI suggestions rely mainly on local code context. 
& Lifecycle-aware context: the system uses relevant requirements, issues, tests, commits, and architectural constraints. 
& Contextual Alignment Score: degree to which a suggestion aligns with project history, conventions, and stated requirements. \\
\midrule
AI outputs are plausible but difficult to verify. 
& Principled explainability: the system provides rationales, assumptions, evidence, and trade-offs. 
& Trust Calibration Time: time and effort required for a developer to verify, accept, revise, or reject a suggestion. \\
\midrule
AI assistance is delivered as a single response. 
& Agentic collaboration: specialized agents coordinate subtasks while preserving human checkpoints. 
& Intervention Quality: whether human review improves the agentic workflow and prevents incorrect or risky changes. \\
\midrule
AI-generated code may pass tests but create future maintenance debt. 
& Long-term maintainability: AI-mediated changes remain understandable and modifiable over time. 
& Maintenance Recovery Cost: effort required for future developers to understand, modify, or repair AI-assisted changes. \\
\midrule
Responsibility for AI-assisted changes is unclear. 
& Accountability and provenance: the system records what was suggested, why, using which evidence, and approved by whom. 
& Traceability Completeness: proportion of AI-assisted changes linked to rationale, source artifacts, and human decisions. \\
\bottomrule
\end{tabular}
\end{table*}

\subsection{Contextual Alignment Score}

We define \textit{Contextual Alignment Score} (CAS) as the degree to which an AI-generated suggestion aligns with the broader project context. This includes alignment with requirements, architectural constraints, coding conventions, dependency policies, historical design decisions, and relevant issue discussions. A high CAS indicates that the suggestion is not merely correct in isolation, but appropriate for the specific software system in which it will be integrated~\cite{jin2024,elsisi2025,caumartin2026}.

CAS could be operationalized using a combination of automated and human measures. Automated signals may include conformity with style rules, test compatibility, dependency constraints, and consistency with similar historical changes. Human evaluation may assess whether the suggestion respects architectural intent, avoids unnecessary complexity, and reflects the rationale expressed in project artifacts~\cite{nascimento2025,dillmann2024}.

\subsection{Trust Calibration Time}

We define \textit{Trust Calibration Time} (TCT) as the time and effort required for a developer to reach an informed judgment about an AI-generated suggestion. This includes the effort needed to understand the change, inspect the explanation, verify assumptions, run tests, compare alternatives, and decide whether to accept, modify, or reject the output~\cite{ahmad2023,patrick2002building,ayele2025developer,Wang2024}.

The goal is not always to minimize TCT. Very low TCT may indicate over-trust if developers accept AI suggestions without adequate review. Very high TCT may indicate that the AI output is difficult to understand or poorly justified. A useful AI teammate should support calibrated trust: developers should be able to quickly accept low-risk, well-explained suggestions while spending more time on uncertain, high-impact, or poorly supported changes~\cite{hamza2024,dubey2020}.

\subsection{An Illustrative Scenario}

To make the framework concrete, we sketch how an AI teammate and its evaluation might operate on a single maintenance task. The scenario is illustrative rather than empirical: the values below are hypothetical and are used to show how the proposed metrics could be computed, not to report measured results.

Consider a regression in a payment service: after a recent change, a subset of refund requests fails with a null-reference error. A developer opens the failing ticket and asks an AI teammate to propose a fix. Rather than editing the open file in isolation, the teammate assembles lifecycle context: the issue report describing the failing refund path, the failing integration test, the commit that last modified the refund handler, the project convention that monetary values use a fixed-point \texttt{Money} type rather than floating point, and the architectural rule that external-facing handlers must validate inputs before dispatch. It proposes a patch that adds the missing validation and explains the change in engineering terms: which commit introduced the regression, which test now protects the behaviour, which convention the fix follows, and one assumption the developer must confirm (that upstream callers never legitimately pass a null payer identifier).

\textbf{Operationalising CAS.} For this task, the Contextual Alignment Score can be computed as a weighted average over a small rubric of project-specific criteria, each scored in $[0,1]$ by an automated check, a human reviewer, or both: conformance to the requirement stated in the ticket; conformance to coding conventions (here, use of the \texttt{Money} type and the validation rule); satisfaction of dependency and API-contract constraints; and consistency with the historically preferred fix for similar regressions. If, for example, a suggestion follows the validation convention and passes the failing test but silently converts \texttt{Money} to a floating-point value, it might score high on requirement conformance and test compatibility yet low on convention conformance, yielding a moderate CAS that correctly flags it as locally correct but contextually misaligned. Reporting the per-criterion breakdown, rather than a single scalar, is what makes CAS diagnostic.

\textbf{Operationalising TCT.} The Trust Calibration Time can be measured by instrumenting the review session and recording the elapsed effort from the moment the suggestion and its explanation are presented to the moment the developer commits to a decision (accept, modify, or reject), optionally decomposed into reading the explanation, inspecting the cited artifacts, running or writing tests, and comparing alternatives. TCT is interpreted against outcomes rather than minimised: a short TCT that ends in accepting a subtly convention-violating patch signals over-trust, whereas a longer TCT spent verifying the null-payer assumption and confirming the fix reflects appropriately calibrated scrutiny. Comparing TCT and decision quality across explanation designs, for instance, with and without the cited commit and test, is one way future studies could evaluate whether an explanation actually improves developer judgement.

This scenario is deliberately small, but it illustrates the shift the framework asks for: the AI teammate is judged not only on whether its patch passes the test, but on whether it used the right lifecycle context, explained itself in terms an engineer can check, and left the developer better able to make an accountable decision.

\subsection{Beyond Correctness}

This framework does not replace traditional measures such as functional correctness, security, performance, or maintainability. Instead, it complements them. A trustworthy AI teammate must still produce technically sound code. However, technical correctness should be evaluated alongside collaborative properties: whether the AI uses appropriate context, explains its reasoning, supports human agency, and leaves behind enough traceability for future maintenance~\cite{siddiq2023,liu2023your,melo2025quality,jamil2025can,dillmann2024}. These collaborative properties should also be assessed across diverse development contexts rather than assumed to transfer from a single setting: teams differ in expertise, infrastructure, and access to AI tools, and a metric that reflects calibrated trust for an expert team may not do so for newcomers~\cite{bohme2025,Pearce2025}.

\section{Research Agenda}

The transition from AI coding assistants to trustworthy AI teammates raises several research challenges for the software engineering community~\cite{lo2023,pezze2025}.

\subsection{RQ1: What Context Should AI Teammates Use?}

More context is not always better. Large amounts of retrieved information can introduce noise, outdated assumptions, privacy risks, and unnecessary cost. Future research should investigate how AI systems can select, prioritize, and explain lifecycle context~\cite{lewis2020,barnett2024,elsisi2025}. This includes studying which artifacts are most useful for different SE tasks, how context quality affects output quality, and how developers interpret AI-selected evidence~\cite{fischer2023,durrani2024}.

\subsection{RQ2: What Explanations Help Developers Make Better Decisions?}

Explainability in SE should be evaluated by its effect on developer judgment. Future studies should examine which explanations help developers detect incorrect suggestions, identify missing assumptions, compare alternatives, and preserve their understanding of the codebase~\cite{ahmad2023,samir2023,agarwal2020}. This requires moving beyond explanation fluency and measuring whether explanations improve human decision-making~\cite{johnson2025explainingsoftwarevulnerabilitieslarge,AlDebeyan2024vulnerability}.

\subsection{RQ3: How Should Agentic Workflows Preserve Human Control?}

Agentic AI systems can decompose complex tasks, but they may also create new forms of opacity. Future research should investigate workflow designs that make agent actions visible, interruptible, and auditable~\cite{he2025,ferrag2025,caumartin2026}. This includes deciding when human approval is required, how intermediate agent outputs should be presented, and how disagreements among agents should be handled~\cite{epperson2025,wang2025}.

\subsection{RQ4: How Does AI-Assisted Code Evolve Over Time?}

Many evaluations of AI-generated code focus on immediate correctness. Software engineering, however, is concerned with evolution. Future research should study whether AI-assisted changes remain understandable and maintainable after weeks, months, or years~\cite{melo2025quality,dillmann2024,jamil2025can}. This includes longitudinal studies of AI-mediated commits, developer comprehension studies, and analyses of whether AI-generated rationale supports future maintenance~\cite{hamza2024,welter2025}.

\subsection{RQ5: How Should Developers Be Trained for Human-AI Collaboration?}

If AI becomes a teammate in software engineering, then developers need new skills. These include prompt formulation, context curation, explanation critique, AI output verification, and responsible use of agentic workflows~\cite{treude2025,alenezi2025,durrani2024}. Software engineering education should prepare students not only to use AI tools, but to question, govern, and collaborate with them~\cite{hamza2024,welter2025}.

\section{Implications}

The human-AI co-evolution agenda has implications for research, practice, and education.

For researchers, it calls for benchmarks that reflect realistic software development settings. Isolated programming problems are not enough. Evaluation should include multi-file systems, evolving requirements, historical artifacts, human review, and maintenance tasks~\cite{chen2021,nascimento2025,dillmann2024}. It should also include diverse development contexts, including teams with different levels of expertise, infrastructure, and access to AI tools~\cite{bohme2025,Pearce2025}.

For practitioners, the agenda suggests that organizations should treat AI-assisted development as a governance issue, not only a productivity issue. Teams need policies for reviewing AI-generated code, recording AI involvement, managing sensitive context, and deciding where agentic automation is appropriate~\cite{ahmad2023,abule2025,zhang2023}. AI adoption should be accompanied by practices that preserve accountability. As a lightweight readiness aid, Table~\ref{tab:swot} frames this decision as a Strengths, Weaknesses, Opportunities, and Threats (SWOT) analysis, pairing the internal strengths and weaknesses of AI-teammate adoption with the external opportunities and threats a team should weigh.

\begin{table}[t]
\centering
\caption{A SWOT view of integrating autonomous AI teammates into a software team, intended as a lightweight readiness aid for practitioners.}
\label{tab:swot}
\footnotesize
\begin{tabular}{p{0.4\textwidth} p{0.4\textwidth}}
\toprule
\textbf{Strengths} & \textbf{Weaknesses} \\
Faster routine implementation; broader language and framework coverage; lower entry barrier; always-available assistance. & Opaque rationale; uneven reliability; limited lifecycle awareness; risk of eroding developer comprehension. \\
\midrule
\textbf{Opportunities} & \textbf{Threats} \\
Lifecycle-aware assistance; better onboarding; wider access across teams and regions; new tooling and evaluation practices. & Over-automation and automation bias; correlated multi-agent failures; unclear accountability; security and maintenance debt. \\
\bottomrule
\end{tabular}
\end{table}

For educators, the agenda suggests that software engineering curricula must evolve. Students should learn how to work with AI assistants critically. This includes understanding their limitations, evaluating generated code, tracing rationale, and maintaining responsibility for final engineering decisions~\cite{treude2025,alenezi2025,durrani2024}. The future developer is not merely a prompt writer, but a software engineer who can collaborate with AI without surrendering judgment.

\section{Conclusion}

AI coding assistants have already changed software engineering practice, but their current form is only an early stage of a larger transformation~\cite{alenezi2025,durrani2024,treude2025}. The next challenge is not simply to generate more code, faster. The challenge is to build AI systems that can participate responsibly in the software lifecycle while preserving human understanding, agency, and accountability~\cite{ahmad2023,bohme2025,lo2023}.

This paper presented a research agenda for trustworthy AI teammates in software engineering. We argued that human-AI co-evolution requires lifecycle-aware context, principled explainability, and agentic collaboration. We also proposed an evaluation framework that moves beyond isolated code correctness toward collaborative quality, including contextual alignment, trust calibration, maintenance recovery, and traceability.

The path from coding tools to AI teammates is not only a technical transition. It is a socio-technical one. If the software engineering community wants AI systems that genuinely improve software quality, productivity, and inclusion, then those systems must be designed for collaboration, not just generation. They must help developers understand what is being changed, why it is being changed, and how those changes will affect the future of the software system.

\bibliographystyle{splncs04}
\bibliography{main}

\end{document}